\documentclass[11pt,a4paper]{article}
\usepackage{jcappub}
\usepackage{bm}
\usepackage{amsfonts}
\usepackage{subfigure}

\newcommand{\be}{\begin{equation}}
\newcommand{\ee}{\end{equation}}
\newcommand{\bea}{\begin{eqnarray}}
\newcommand{\eea}{\end{eqnarray}}

\renewcommand{\vec}[1]{{\bm #1}}
\newcommand{\den}{{\rm g}\,{\rm cm}^{-3} }
\newcommand{\meff}{m^{\rm \star} }
\newcommand{\Msun}{M_{\odot}}

\DeclareMathAlphabet\mathrsfso {U}{rsfso}{m}{n}

\begin{document}

\title{Revisiting the SN1987A gamma-ray limit on ultralight axion-like particles}

\author[a]{Alexandre~Payez}
\author[b]{Carmelo~Evoli}
\author[c]{Tobias~Fischer}
\author[d]{Maurizio~Giannotti}
\author[b]{Alessandro~Mirizzi}
\author[a]{Andreas~Ringwald}

\affiliation[a]{Theory group, Deutsches Elektronen-Synchrotron DESY\\Notkestra\ss{}e 85, D-22607 Hamburg, Germany}
\affiliation[b]{II.~Institut f\"ur Theoretische Physik, Universit\"at
Hamburg\\ Luruper Chaussee 149, D-22761 Hamburg, Germany}
\affiliation[c]{Institute for Theoretical Physics, University of Wroc\l{}aw
\\Pl. M. Borna 9, 50-204 Wroc\l{}aw, Poland} 
\affiliation[d]{Physical Sciences, Barry University\\
11300 NE 2nd Ave., Miami Shores, FL 33161, USA}

\emailAdd{alexandre.payez@desy.de}
\emailAdd{carmelo.evoli@desy.de}
\emailAdd{fischer@ift.uni.wroc.pl}
\emailAdd{mgiannotti@barry.edu}
\emailAdd{alessandro.mirizzi@desy.de}
\emailAdd{andreas.ringwald@desy.de}

\abstract{%
We revise the bound from the supernova SN1987A on the coupling of ultralight axion-like particles (ALPs) to photons. In a core-collapse supernova, ALPs would be emitted via the Primakoff process, and eventually convert into gamma rays in the magnetic field of the Milky Way. The lack of a gamma-ray signal in the GRS instrument of the SMM satellite in coincidence with the observation of the neutrinos emitted from SN1987A therefore provides a strong bound on their coupling to photons. Due to the large uncertainty associated with the current bound, we revise this argument, based on state-of-the-art physical inputs both for the supernova models and for the Milky-Way magnetic field. Furthermore, we provide major amendments, such as the consistent treatment of nucleon-degeneracy effects and of the reduction of the nuclear masses in the hot and dense nuclear medium of the supernova. With these improvements, we obtain a new upper limit on the photon--ALP coupling:
 \begin{equation}
 g_{a\gamma} \lesssim 5.3 \times 10^{-12}\,\ \textrm{GeV}^{-1},\,\  \,\ \textrm{for}\,\ \,\ m_{a} \lesssim 4.4 \times 10^{-10}~\textrm{eV} \,\ , \notag
 \end{equation}
and we also give its dependence at larger ALP masses $m_a$.
Moreover, we discuss how much the Fermi-LAT satellite experiment could improve this bound, should a close-enough supernova explode in the near future.
}
\maketitle

\section{Introduction}

Axion-like particles (ALPs) are very light pseudoscalar bosons $a$, with a two-photon coupling 
described by the Lagrangian
\begin{equation}
{\mathcal L}_{a\gamma}= -\frac{1}{4} g_{a\gamma} F_{\mu\nu}{\tilde F}^{\mu \nu} a = g_{a\gamma}\,
{\bf E}\cdot {\bf B}\, a \,\ ,
\label{eq:axlagrang}
\end{equation}
where ${\tilde F}^{\mu \nu}$ is the dual of the electromagnetic field $F_{\mu\nu}$, and $g_{a\gamma}$ is the photon--ALP coupling constant with dimensions of inverse energy. ALPs are predicted by several extensions of the Standard Model, such as four-dimensional models (see e.g.\@ Refs.~\cite{Anselm:1981aw,susy1,Dias:2014osa}), Kaluza--Klein theories (see e.g.\@ Ref.~\cite{kaluzaklein1}), and especially superstring theories (see e.g.\@ Refs.~\cite{witten3,Acharya:2010zx,witten4}); for a review, see e.g.\@ Refs.~\cite{masso,Jaeckel:2010ni,Ringwald:2014vqa}. 

In the presence of an external magnetic field ${\bf B}$, the $a \gamma \gamma$ coupling entails that interaction eigenstates differ from propagation eigenstates, thereby leading to the phenomenon of photon--ALP conversion $\gamma \leftrightarrow a$, and in particular to photon--ALP oscillations~\cite{sikivie,Raffelt:1987im,Anselm:1987vj}. This mixing effect is exploited to search for generic ALPs in light-shining-through-the-wall experiments (e.g.\@  ALPS~\cite{Ehret:2010mh}, CROWS~\cite{Betz:2013dza}, and OSQAR~\cite{Ballou:2014rra}), and for ALP dark matter~\cite{Arias:2012az} in micro-wave cavity experiments (for instance, ADMX~\cite{Duffy:2006aa}). Photon--ALP oscillations would also lead  to peculiar signatures in astrophysical and
cosmological observations~\cite{Jaeckel:2010ni,Mirizzi:2007hr,Dupays:2005xs,Payez:2011sh}. In particular, an intriguing hint for these particles has been recently suggested by very-high-energy gamma-ray experiments: in fact, photon--ALP  conversions in large-scale cosmic magnetic fields would modify the opacity of the universe to TeV photons and this might explain an anomalous spectral hardening in very-high-energy gamma-ray spectra~\cite{De Angelis:2007yu,HS:2007,Horns:2012fx,Horns:2012kw,Rubtsov:2014uga,Tavecchio:2014yoa}. Moreover, if they existed, similar ALPs might also  explain the presence of a soft X-ray excess reported in galaxy clusters~\cite{Conlon:2013txa,Angus:2013sua}.

Several new experiments and astrophysical considerations have helped probing large regions of the ALP parameter space; for a recent review, see e.g.\@ Ref.~\cite{Carosi:2013rla}. In particular, the $g_{a\gamma}$ vertex would allow for the production of ALPs via the Primakoff process in the fluctuating electric fields of nuclei and electrons in a stellar plasma~\cite{Raffelt:1985nk}. 
The predicted Solar ALP spectrum is currently searched for by the CERN Axion Solar Telescope (CAST)~\cite{Andriamonje:2007ew}, looking for conversions of Solar axions and ALPs into X-rays in a dipole magnet tracking the Sun. The CAST experiment has achieved the best experimental bound on the photon--ALP coupling, obtaining $g_{a\gamma}\lesssim 8.8\times 10^{-11}$~GeV$^{-1}$ for ALP masses $m_a\lesssim 0.02$~eV~\cite{Arik:2011rx}. ALP production in stars via the Primakoff process would also cause an additional energy drain that may change the stellar lifetime, eventually beyond the limits allowed by astronomical observations~\cite{Raffelt:1985nk,Raffelt:1987yu,Raffelt:2006cw,Friedland:2012hj,Ayala:2014pea}. In particular, one finds $g_{a\gamma}\lesssim 8\times 10^{-11}$~GeV$^{-1}$ from Cepheid stars~\cite{Friedland:2012hj}, and $g_{a\gamma}\lesssim\nobreak 6.6\times \nobreak10^{-11}$~GeV$^{-1}$ from globular cluster stars~\cite{Ayala:2014pea} (see also Ref.~\cite{Raffelt:1987yu}), superseding the CAST direct bound.
At much smaller masses, further limits have for instance been derived from magnetic white dwarfs~\cite{Gill:2011yp}, and from the non-observation of irregularities in TeV photon spectra~\cite{Abramowski:2013oea}.
Finally, ultralight ALPs mixing with photons in external magnetic fields like that of the Virgo supercluster plane would affect the linear and circular polarizations of light from quasars. Assuming that this magnetic field, extended over $\sim5$~Mpc, can be decomposed in 100-kpc domains, with an average electron density $n_{e,0}$ between $10^{-6}$ and $10^{-5}$~cm$^{-3}$, the limit scales as $g_{a\gamma} \lesssim 6.3 \times 10^{-12}~{\rm GeV}^{-1} \frac{2~\mu{\rm G}}{|\vec{B}_{\rm \scalebox{0.6}{domain}}|} \left(\frac{n_{e,0}}{10^{\scalebox{0.5}{$-5$}}~{\rm cm}^{\scalebox{0.5}{$-3$}}}\right)^{\scalebox{0.65}{$1.3$}}$ for $m_a \lesssim 4 \times 10^{-14}$~eV~\cite{Payez:2012vf,Payez:2013yxa}. 

For ALPs with masses $m_a \lesssim 10^{-9}$~eV, the strongest bound on $g_{a\gamma}$ comes  from the absence of gamma rays from SN1987A, which exploded in the Large Magellanic Cloud at a distance of 50~kpc as a core-collapse supernova.
ALPs produced inside a supernova (SN) core via the Primakoff process with energies $E\sim 100$~MeV would escape essentially freely and the emitted SN ALP flux could then be converted into photons in the magnetic field of the Milky Way, leading to a gamma-ray flux. The latter would be observable hours before the optical flash associated with the SN explosion reaching the stellar surface. At the time at which the neutrinos from SN1987A were observed, the Gamma-Ray Spectrometer (GRS) of the Solar Maximum Mission (SMM) was operative and could have potentially detected the signal of this ALP conversion (see Ref.~\cite{Raffelt:1996wa} for a review). However, there has been no evidence for such a gamma-ray excess from that SN in coincidence with the neutrino observation. This was therefore used to strongly constrain the ALP--photon coupling, namely  $g_{a\gamma}\lesssim 1 \times 10^{-11}$~GeV$^{-1}$~\cite{Brockway:1996yr} or even $g_{a\gamma} \lesssim 3\times 10^{-12}$~GeV$^{-1}$~\cite{Grifols:1996id}. These long-standing bounds have received renewed attention in the last few years. Indeed, the astrophysical hints from TeV photons require ALPs with $m_a \lesssim 10^{-7}$~eV and $g_{a\gamma}\gtrsim 10^{-12}\textrm{--}10^{-11}$~GeV$^{-1}$ depending on the assumptions for the cosmological magnetic field~\cite{Meyer:2013pny}. Moreover, this range is within the reach of the planned upgrade of the photon-regeneration experiment ALPS at DESY~\cite{Bahre:2013ywa}, and of the next-generation Solar-ALP detector IAXO (International Axion Observatory)~\cite{Irastorza:2011gs,Armengaud:2014gea}. Therefore, the SN1987A bound is now crucial to corner the parameter space available for ultralight ALP searches.

The bounds quoted in Refs.~\cite{Brockway:1996yr,Grifols:1996id} are however affected by large uncertainties. In particular, the treatment of the ALP production in the SN core and of the ALP--photon conversions in the Milky Way are rather schematic in those papers. Hence, the SN1987A limit has been questioned in the recent literature (especially its ALP-mass dependence) and occasionally even objected~\cite{HS:2007,Hooper:2007bq,Csaki:2001yk, De Angelis:2007yu,SanchezConde:2009wu}.

Given the relevance of a robust constraint for the current ALP searches, after almost twenty years since the original analyses, the general focus of this article is on the re-evaluation of the SN1987A limit with state-of-the-art physics inputs. In particular, we aim at amending and restating different aspects of the previous analyses: \emph{(a)} we calculate the SN ALP flux using the results from recent long-term simulations of core-collapse SN explosions~\cite{Fischer:2009af,Fischer:2012}, which are based on general-relativistic radiation hydrodynamics with three-flavor Boltzmann neutrino transport in spherical symmetry; \emph{(b)} we apply a nuclear microscopic description of the SN plasma, i.e.\@ including effects of proton degeneracy and of the reduction of the proton mass in the dense stellar medium; \emph{(c)} we accurately characterize the ALP--photon conversions in the Milky Way, solving the oscillation equations using sophisticated models of the Galactic magnetic field~\cite{Jansson:2012pc,Pshirkov:2011um}. As a result of our analysis, we find a new upper limit:
\begin{equation}
g_{a\gamma} \lesssim 5.3 \times 10^{-12}\,\ \textrm{GeV}^{-1},\,\  \,\ \textrm{for}\,\ \,\ m_{a} \lesssim 4.4 \times 10^{-10}~\textrm{eV} \,\ .\nonumber
\end{equation}

The manuscript is organized as follows. In Sec.~\ref{sec:ALPprod} we discuss the ALP production via the Primakoff process inside the SN core. In Sec.~\ref{sec:conv} we describe the ALP--photon conversions in the Milky-Way magnetic field. In Sec.~\ref{sec:results} we present our new upper limit from the bound on gamma rays obtained by the Gamma-Ray Spectrometer during the SN1987A explosion. Finally, in Sec.~\ref{sec:ccl} we summarize our results and conclude. We also comment on the Fermi-LAT satellite experiment capability to probe this coupling, should a sufficiently close SN explode in the near future. 

\section{ALP production in a supernova core} \label{sec:ALPprod} 
  
\subsection{Core-collapse supernova model}
\label{sec:SN}
 
In order to calculate the SN ALP flux, we consider in this study the core-collapse supernova simulations of massive progenitor stars with $10.8$ and $18.0$~$M_{\odot}$ from Ref.~\cite{Fischer:2009af}. These iron-core progenitors have been evolved consistently through all supernova phases up to several tens of seconds after the onset of the supernova explosion, using three-flavor Boltzmann neutrino transport within the spherically symmetric and general-relativistic radiation-hydrodynamics framework. Since explosions cannot be obtained in spherical symmetry for such massive stellar models, the neutrino energy deposition had been enhanced; for details, see Refs.~\cite{Fischer:2009af,Fischer:2012}. In this study we focus on the long-term signal of ALP emission, on a timescale of the order of tens of seconds after the supernova explosion has been launched. The SN explosion is driven via energy liberation from the SN core onto its surface layer in a highly turbulent hydrodynamics environment. It takes place on a timescale of only several 100 milliseconds leading to the ejection of the stellar mantle, which is still highly uncertain and subject of active research. On the other hand, the physics of the epoch after the supernova explosion has been launched is rather well under control (cf. Refs.~\cite{MartinezPinedo:2012,Roberts:2012}). Moreover, this phase of the supernova evolution is moderately independent from details of the explosion mechanism. It can also well be simulated in a spherically symmetric setup since multi-dimensional phenomena such as convection, rotation and magnetic fields play a minor role. Furthermore, it also refers to the epoch during which most of the energy (of $\mathcal{O}(10^{53}~\textrm{erg})$) is released via the continuous emission of neutrinos of all flavors, which drives the evolution towards the final cold and neutrino-less neutron star. Comparing the energy release from ALP emission ($\lesssim 10^{50}~\textrm{erg}\times {(g_{a\gamma}/10^{-10}~{\rm GeV}^{-1})}^2$) to this neutrino energy loss allows us to neglect its feedback on the supernova evolution.

For the current study of ALP production, we are interested in the deep interior of the nascent protoneutron star (PNS) and in its dynamical evolution. The PNS forms when the collapsing stellar core reaches central densities in excess of saturation density, and consequently bounces back. Being still hot and lepton rich, the PNS cools via the emission of neutrinos of all flavors on a timescale of the order of tens of seconds after the SN explosion has been launched. The state of matter inside the PNS\hspace{1pt}---\hspace{1pt}densities in excess of nuclear saturation density ($\rho_0=2.6\times10^{14}$~g~cm$^{-3}$) and temperatures $T$ of several tens of MeV\hspace{1pt}---\hspace{1pt}has a large neutron excess determined by the electron fraction $Y_e\simeq0.05$\textrm{--}$0.2$.
In general, the contraction of the PNS depends on two major aspects: the compression behavior of the high-density nuclear equation of state (EOS), and the deleptonization rate given by the diffusion of neutrinos from the PNS interior towards the surface where they decouple from matter.

\begin{figure}[!t]
\centering
\subfigure[Temperature in units of MeV]{
\includegraphics[width=0.475\columnwidth]{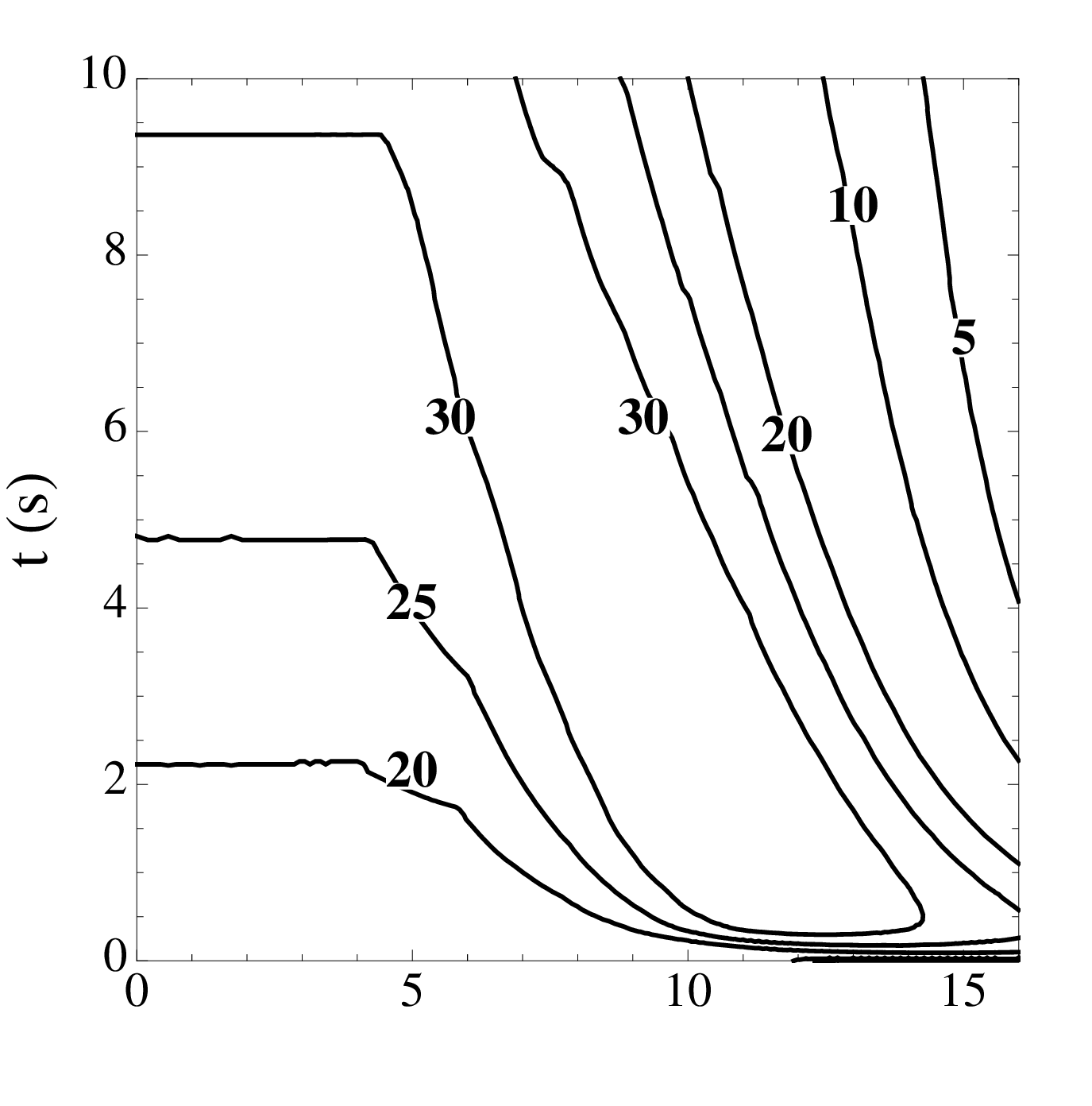}
\label{fig:contour_T}}
\hfill
\subfigure[Density in units of $10^{14}$~g~cm$^{-3}$]{
\includegraphics[width=0.47555\columnwidth]{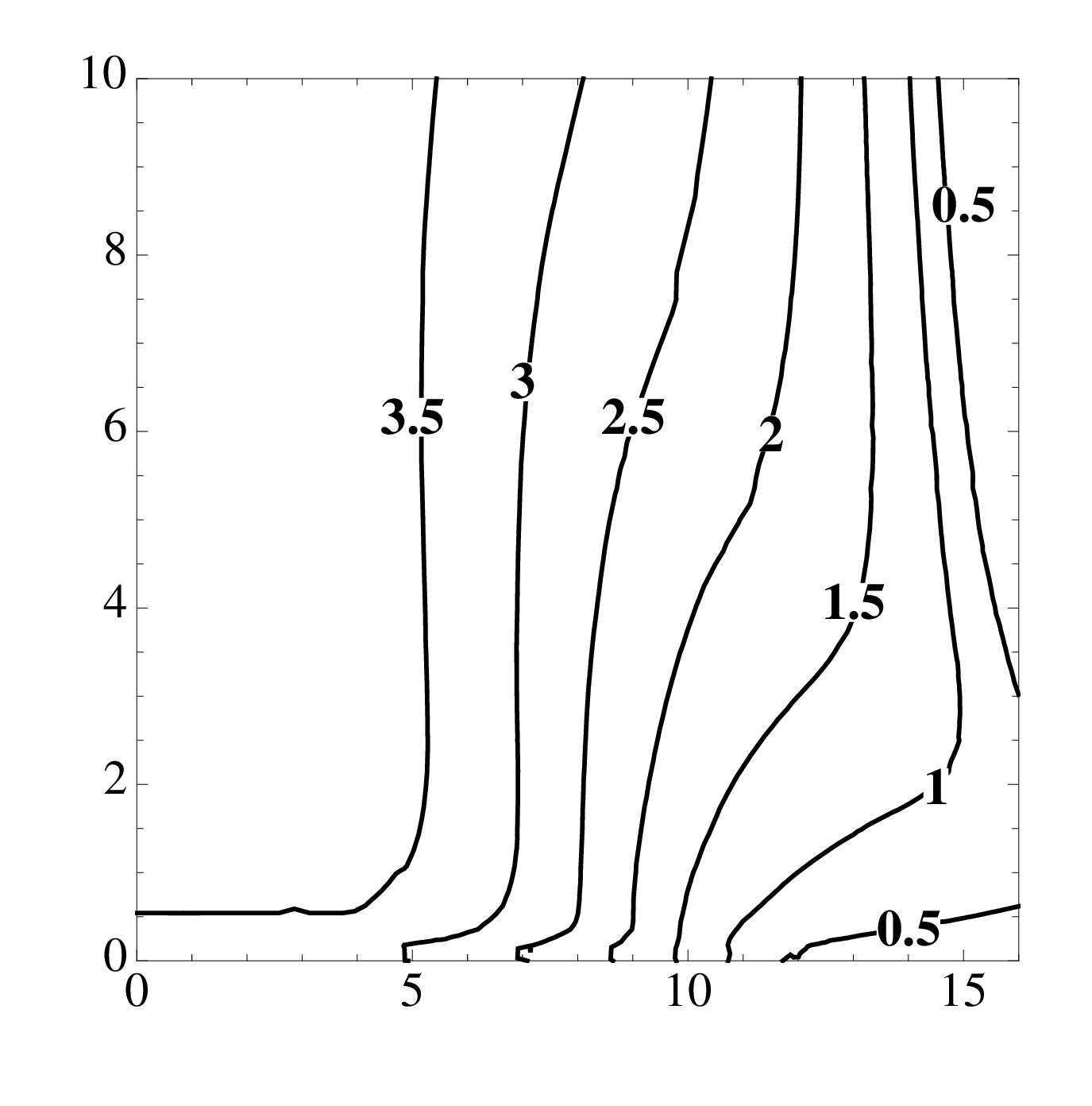}
\label{fig:contour_d}}
\\
\subfigure[Proton mass in units of MeV]{
\includegraphics[width=0.475\columnwidth]{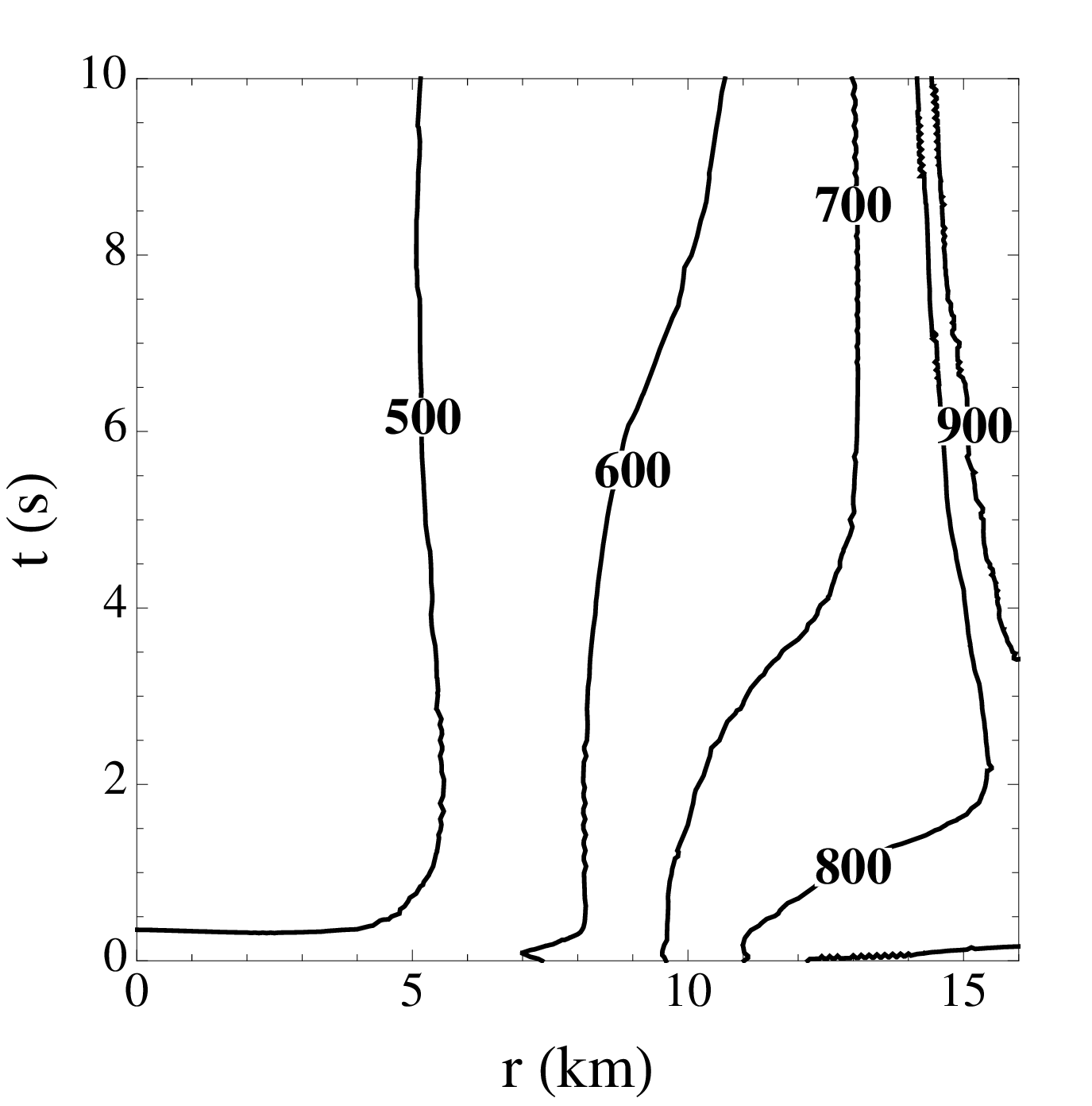}
\label{fig:contour_mp}}
\hfill
\subfigure[Proton degeneracy, $\eta_p=(\mu_p-\meff_p)/T$]{
\includegraphics[width=0.47555\columnwidth]{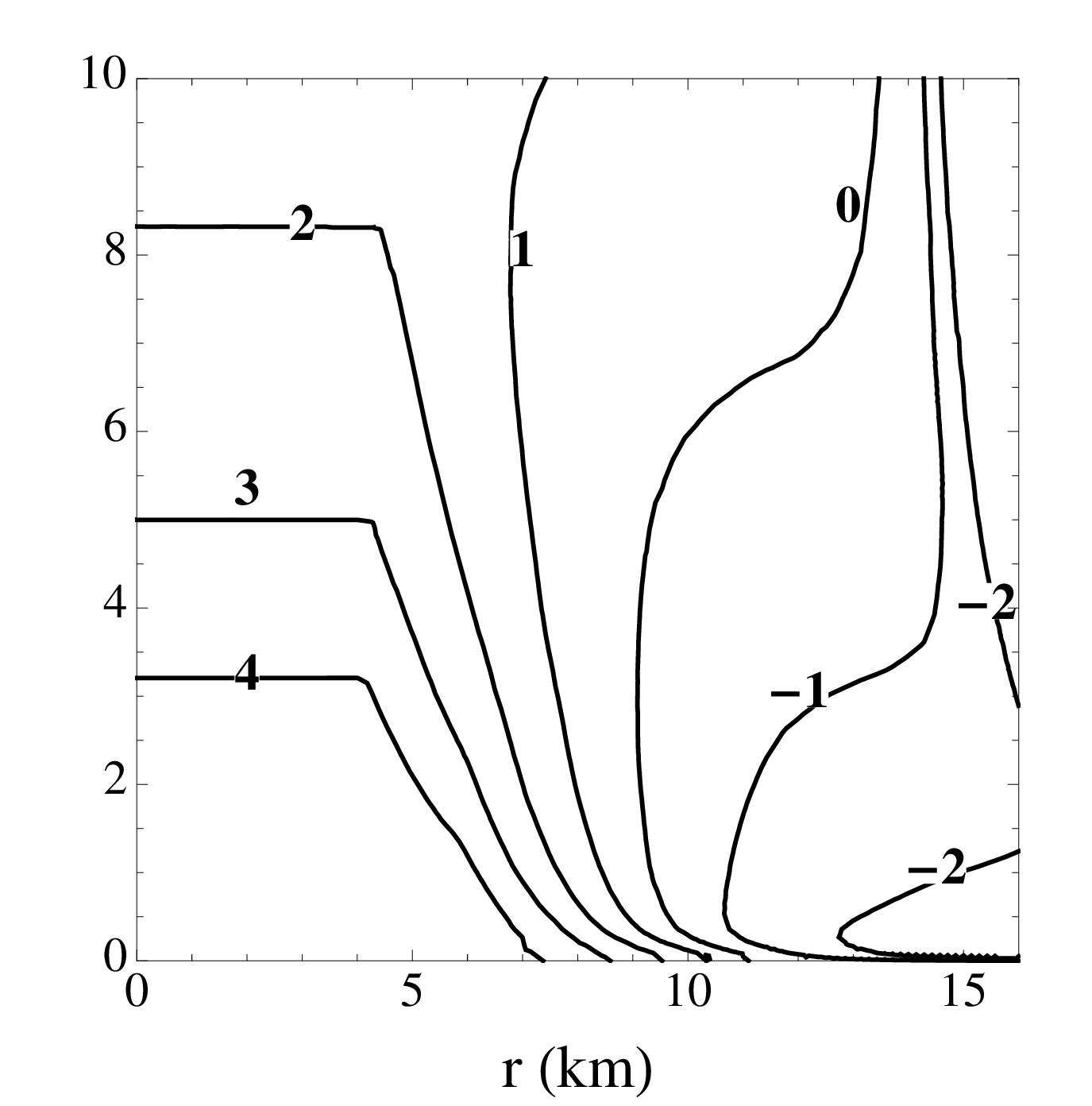}
\label{fig:contour_eta}}
\caption{Contours of selected quantities showing the evolution during the first 10~s after the core bounce of the inner most 16~km of the central PNS for the simulation using the 18~M$_\odot$ progenitor star based on the nuclear EOS from Ref.~\cite{Shen:1998}. }
\label{fig:contours}
\end{figure}

Henceforth, we select the simulation of the 18~$M_\odot$ progenitor as our reference model, which will be further discussed in the following paragraphs; the progenitor of SN1987A has indeed been identified as a massive star with $20\pm2$~M$_\odot$~\cite{Shigeyama:1990,Blinikov:2000,Utrobin:2011}. Figures~\ref{fig:contour_T} and~\ref{fig:contour_d} show contours for the relevant temperature and density evolutions, illustrating the PNS contraction behavior which is typically realized in all simulations of the deleptonization phase after the explosion onset. We see a slow rise of the core density and of the temperature, reaching densities of several times $10^{14}$~g~cm$^{-3}$ and temperatures of $20$\textrm{--}$40$~MeV, on a timescale of about 10~s. Note that at early times after the core bounce, $t\simeq1$\textrm{--}$3$~s, the highest temperatures are reached several km off-center, which is related to the non-monotonous temperature profiles due to the supernova history prior to the explosion onset. Only at late times ($t\gtrsim10$~s) does the temperature decrease monotonously from center towards the PNS surface with maximum at the very center.

Relevant for the present study, in particular for the ALP production rate, are two nuclear matter aspects: the reduction of the nuclear masses due to medium effects (shown in Fig.~\ref{fig:contour_mp}), and the possible degeneracy of protons (Fig.~\ref{fig:contour_eta}). Both phenomena depend on the nuclear physics model, for which we apply here the same nuclear EOS that was used in the supernova simulations~\cite{Shen:1998}. In this relativistic mean-field treatment, the reduction of the nucleon masses due to the scalar interactions at high density leads to effective masses $\meff_N$, which replace the vacuum masses $m_N$ in all the expressions used to determine the microscopic processes, like weak reactions and the production of ALPs. Figure~\ref{fig:contour_mp} shows that this reduction can be substantial, $\meff_N/m_N\simeq0.5$, at densities in excess of saturation density $\rho_0$. In addition, the proton degeneracy $\eta_p = (\mu_p - \meff_p)/T$ can be large, i.e.\@ $\eta_p>1$ (see Fig.~\ref{fig:contour_eta}). The role of these effects related to nuclear physics on the production of ALPs is discussed further below.

\subsection{Primakoff rate}  

ALPs coupled to the electromagnetic radiation as in Eq.~\eqref{eq:axlagrang} are produced in the stellar medium primarily through the Primakoff  process~\cite{Raffelt:1985nk}, in which thermal photons are converted into ALPs in the electrostatic field of ions, electrons and protons.

Using the Heaviside--Lorentz convention for electromagnetism, the Primakoff conversion rate per unit time of photons into pseudoscalars is given by the following expression,
\begin{equation}
\Gamma= \frac{g_{a\gamma}^2\, \alpha \, n_p^{\rm eff}}{8}
\left[\bigg(1+\frac{\kappa^2}{4 E^2}\bigg)\ln\bigg(1+\frac{4 E^2}{\kappa^2}\bigg) -1\right] \,,
\label{eq:primakoff}
\end{equation}
where $ \alpha$ is the electromagnetic fine-structure constant, $ n_p^{\rm eff} $ the effective number of targets, $E$ the photon energy, and $\kappa$ an appropriate screening scale which accounts for the finite range of the electric field of the charged particles in the stellar medium. This rate had been derived assuming that photons can be treated as being massless particles~\cite{Raffelt:1985nk}, which is essentially the case inside a SN core~\cite{Brockway:1996yr}. It can be obtained from the expression of the plasma frequency in the relativistic limit, which is also valid in the degenerate case (see e.g.\@ Ref.~\cite{Raffelt:1996wa}).\footnote{Formally speaking, inside of a plasma the photon dispersion relation becomes $\sqrt{\omega^2 - {\omega_{\rm pl}}^2}$, where the plasma frequency $\omega_{\rm pl}$ acts as an effective photon mass (it is also the dominant contribution inside of a SN~\cite{Kopf:1997mv}).}

As we have already discussed and as shown in Fig.~\ref{fig:contours}, matter inside the PNS is characterized by temperatures of a few 10 MeV and densities of  the order of $ 10^{14}~\den$ during the first few seconds after the explosion. Under such conditions, electrons are highly degenerate: their phase space is Pauli-blocked and hence their contribution to the ALP production is negligible. On the other hand, protons are only partially degenerate (cf.~Fig.~\ref{fig:contour_eta}). They can contribute more substantially to the emission rate. Hence, we will follow Refs.~\cite{Brockway:1996yr,Grifols:1996id} and ignore electron contribution to the Primakoff rate\footnote{The Primakoff rate given in Eq.~\eqref{eq:primakoff} has been derived for negligible photon energies compared to the electron mass, for which the interferences with the Compton production on electrons can be neglected~\cite{Raffelt:1985nk}. This simplification remains valid for the present analysis, since electrons are not available as targets and since the proton mass is much higher than the photon energies. Moreover rate \eqref{eq:primakoff} can of course be used should we decide to consider ALPs only interacting with photons via the Lagrangian~\eqref{eq:axlagrang}.}. Consequently, the effective number of targets has been indicated as $ n_p^{\rm eff} $ in Eq.~\eqref{eq:primakoff}. Note that previous analyses ignored the proton degeneracy on the account that the proton Fermi energy is never much higher than the temperature, due to the large proton rest mass. Therefore, protons are likely to be at most partially degenerate. In the present study of ALP production, for the first time we properly quantify this aspect consistently based on a microscopic description of nuclear matter. Moreover, as discussed above, medium modifications in the hot and dense stellar matter reduce the nuclear masses (see Fig.~\ref{fig:contour_mp}), which in turn enhances the proton degeneracy.

Contours of the proton degeneracy parameter, $ \eta_p=(\mu_p-\meff_p)/T $, where $ \mu_p$ is the proton chemical potential and $\meff_p$ the proton effective mass in medium, are shown in Fig.~\ref{fig:contour_eta}. The $ \eta_p $ parameter is implicitly defined through the equation for the proton number density,
\begin{eqnarray}
\label{Eq:np}
n_{p} &=& 2\int \dfrac{d^3p}{(2\pi)^3}\,\left(\exp{\left( \frac{E_p(\vec{p})-\mu_p}{T} \right)} +1 \right)^{-1}\nonumber
	\\
      &=& 2\int \dfrac{d^3p}{(2\pi)^3}\,\left(\exp{\left( \frac{p^2}{2\meff_p T} - \eta_p \right)} +1 \right)^{-1} \,\,\,,
\end{eqnarray}
where we have used the non-relativistic dispersion relation, $E_p (\vec{p})=p^2/(2\meff_p) + \meff_p$. Expression~\eqref{Eq:np} characterizes the degeneracy of the nucleons in the stellar interior. Protons are essentially degenerate for $\eta_p$ greater than unity and non-degenerate for $ \eta_p<0 $. As evident from the figure, the non-degenerate approximation treatment of protons in the stellar medium is generally not justified and hence in this work we will not make this assumption. 

\begin{figure}[!t]
\centering
\subfigure[Effective proton mass over temperature, $ \meff/T $]{
\includegraphics[width=0.475\columnwidth]{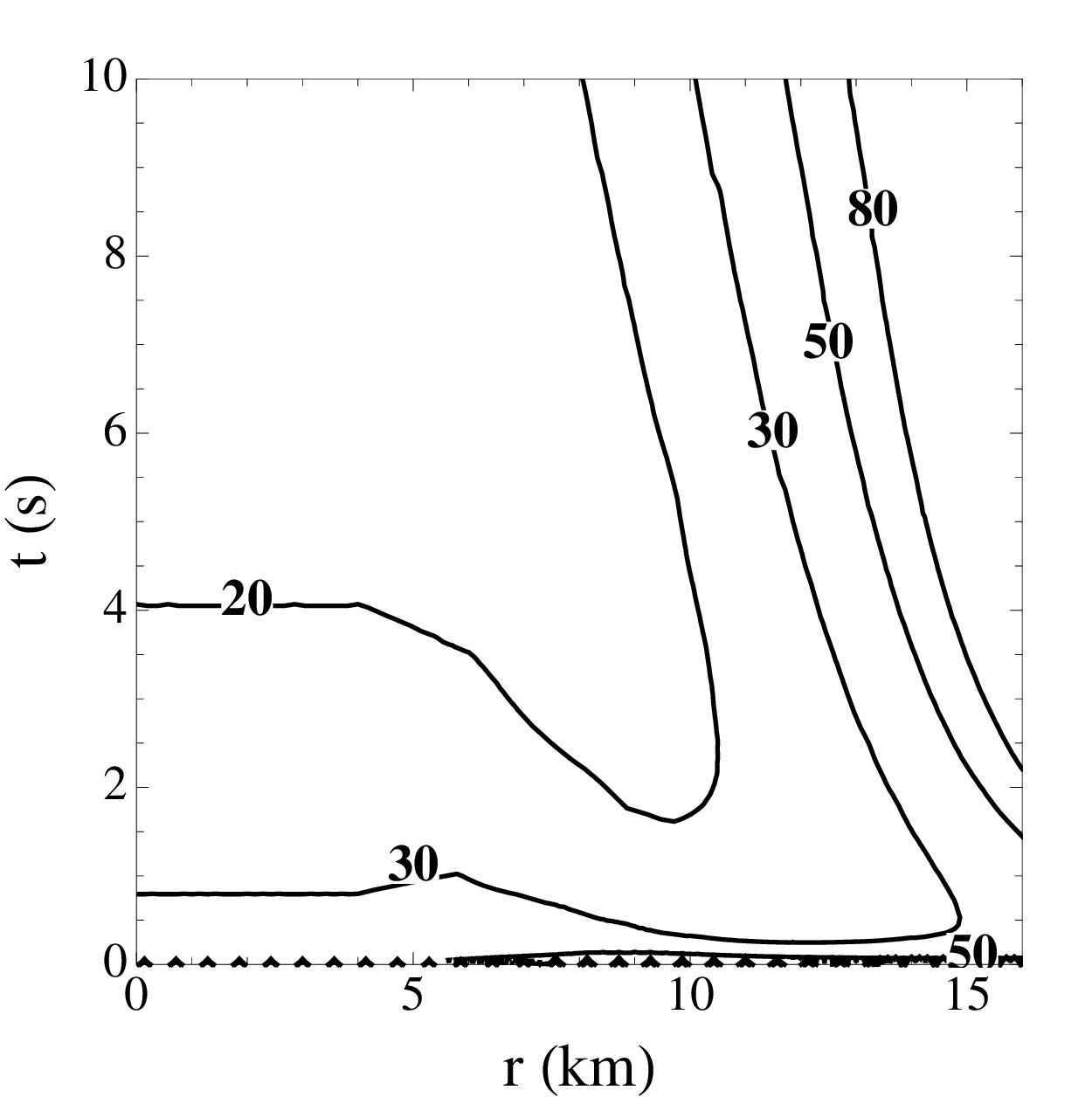}
\label{fig:contour_mp_over_T}}
\hfill
\subfigure[$ n_p^{\rm eff}/n_p $, as defined in Eq.~\eqref{Eq:Rdeg}]{
\includegraphics[width=0.47555\columnwidth]{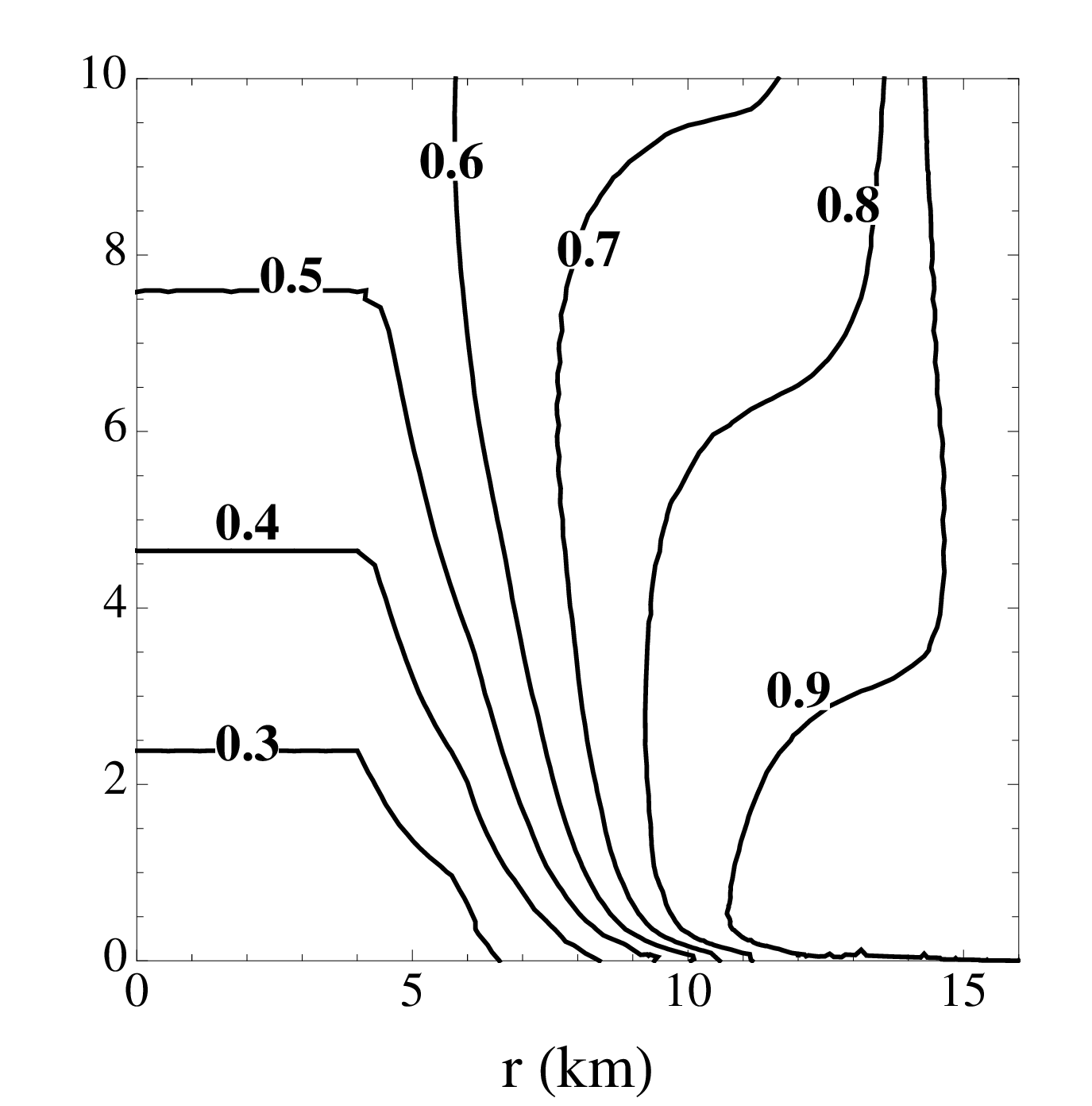}
\label{fig:contour_Rdeg}}
\caption{Same as Fig.~\ref{fig:contours}, but for different quantities.}
\label{fig:contours_deg}
\end{figure}

The degeneracy of the targets complicates considerably the calculation of the Primakoff emission rate. However, a few considerations will help simplify our analysis. First of all, we can easily estimate the reduction of the number of targets, $n_p\rightarrow n_p^{\rm eff} $, if we ignore the proton recoil. This premise is already implicitly given in Eq.~\eqref{eq:primakoff}. It is well justified even when we account for the reduced proton mass in the medium. Indeed, our numerical analysis shows that the temperature is always much smaller than the effective nucleon mass~(cf. Fig.~\ref{fig:contour_mp_over_T}). With this assumption, the effective number of targets can be calculated as follows,
\begin{eqnarray}
\label{Eq:Rdeg}
\dfrac{n_p^{\rm eff}}{n_p}=\dfrac{2}{n_p}\int \dfrac{d^3p}{(2\pi)^3}f_p(1-f_p)\,,
\end{eqnarray}
where $f_p $ is the Fermi--Dirac distribution function for the proton (see, e.g.\@ Ref.~\cite{Raffelt:1996wa}). Contours of $n_p^{\rm eff}/n_p$ are shown in Fig.~\ref{fig:contour_Rdeg} and indicate a suppression of more than 50\%  close to the very dense stellar center.

\begin{figure}
\centering
\includegraphics[width=.5\columnwidth]{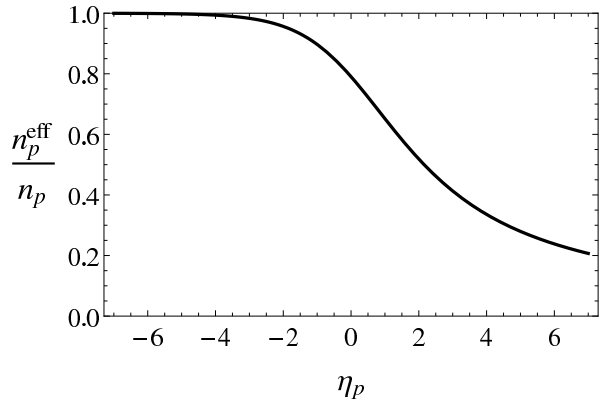}
\caption{Reduction of the number of targets, as obtained from Eq.~\eqref{Eq:Rdeg} for different values of the proton degeneracy parameter $\eta_p$. The same mathematical function also enables a description of the screening length between the Debye and Thomas--Fermi regimes (see main text).}
\label{fig:Rdeg}
\end{figure}

Another difficulty is the appropriate choice of the screening length in Eq.~\eqref{eq:primakoff}. The Debye scale,
\begin{eqnarray}
\label{Eq:kappa_Debye}
\kappa_{\rm Debye}^2=\dfrac{4\pi\alpha n_p}{T} \,,
\end{eqnarray}
is appropriate in the non-degenerate regime, while in the degenerate case the Thomas--Fermi scale should be used. In general, the screening length for the Coulomb potential in a weakly coupled plasma is controlled by the longitudinal component of the polarization tensor (see, e.g.\@ Ref.~\cite{Raffelt:1996wa}). Using the one-loop, non-relativistic approximation for the polarization tensor, we find
\begin{eqnarray}
\kappa^2\simeq \dfrac{4\alpha\meff_p}{\pi}\int dp f_p.
\end{eqnarray}
In the non-degenerate limit this is just the Debye screening length (\ref{Eq:kappa_Debye}) while in the degenerate limit it becomes the Thomas--Fermi scale. An integration by parts leads to $\kappa^2/ \kappa_{\rm Debye}^2= n_p^{\rm eff}/n_p$, 
showing that the reduction of the targets and the modification of the screening length are controlled by exactly the same function, illustrated in Fig.~\ref{fig:Rdeg}. Therefore, in our analysis we will use the screening scale as follows,
\begin{eqnarray}
\label{Eq:kappa}
\kappa^2 = \kappa_{\rm Debye}^2 \frac{n_p^{\rm eff}}{n_p}
= \dfrac{4\pi\alpha n_p^{\rm eff}}{T} \,.
\end{eqnarray}
From Fig.~\ref{fig:contour_Rdeg}, we see that the effect of the degeneracy on the screening length can be substantial inside the SN core.

With these assumptions, the ALP volume production rate per unit energy can be calculated by multiplying Eq.~\eqref{eq:primakoff} with the density of thermal photons and accounting for the reduction of the number of targets. We find
\begin{eqnarray}
\dfrac{d \dot n_a}{dE}=
\frac{g_{a\gamma}^{2}\xi^2\, T^3\,E^2}{8\pi^3\, \left( e^{E/T}-1\right) }
\left[ \left( 1+\dfrac{\xi^2 T^2}{E^2}\right)  \ln(1+E^2/\xi^2T^2) -1 \right] \,,
\label{eq:axprod}
\end{eqnarray}
where 
\begin{eqnarray}
\xi^2=\dfrac{\kappa^2}{4T^2}\,.
\end{eqnarray}
Notice that this equation is formally identical to the one used in Ref.~\cite{Brockway:1996yr}; all our changes are included in the definition of the screening length. Note further that we neglect contributions from protons bound in nuclei for the ALP production rate; in particular light clusters but also heavy clusters become abundant at low temperatures and densities~\cite{Fischer:2014}. However, from Eq.~\eqref{eq:axprod} we expect that the dominant ALP production is associated with high temperatures where nuclei cannot exist. We leave the aspect of nuclear clusters for future explorations.

\subsection{Supernova ALP flux} 

We now turn to the actual calculation of the production of ALPs inside a given SN, as it evolves right after the core bounce. We closely follow the steps given in Ref.~\cite{Brockway:1996yr}, while using the improved Primakoff rate and state-of-the-art SN simulations. These numerical results consist of 658 (resp.\@~592) snapshots describing the evolution of the various physical quantities entering the volume ALP production rate for a $18$~M$_{\odot}$ (resp.\@~$10.8$~M$_{\odot}$) progenitor, during 21.8~s (resp.\@~10.5~s) after the core bounce.

First of all, we want to calculate the total ALP production rate per unit energy:
\begin{equation}
\frac{d\dot{N}_a}{dE} = \int
\frac{d\dot{n}_a}{dE}\ d^3r\,\,\,,
\end{equation}
by integrating Eq.~\eqref{eq:axprod} over the volume of the supernova. The corresponding integration limits and $dr$ are defined via the discretized numerical radial grid used in the SN simulations, for which an adaptive mesh-refining mass grid has been used.
In fact we only consider the contributions to the ALP production from radii up to $r_{\rm max} = 50$~km. We do this for consistency: to remain safely in the validity domain of the EOS of Ref~\cite{Shen:1998}, used both to include the mass-reduction effects and in the SN simulations themselves, which requires the temperature to be higher than 0.45~MeV.\footnote{For the $10.8$~$M_{\odot}$ progenitor this is verified until $t=10.5$~s for all radii below 50 km, while for the $18$~$M_{\odot}$ progenitor it is until $t\sim 18$~s. For our bound, we do not need to consider later times than any of these.} Again since the Primakoff production scales mainly with the temperature, the ALP production from larger radii (hence lower temperatures) has a negligible contribution and can well be ignored. 

\begin{figure}
\centering
\includegraphics[width=.7\columnwidth]{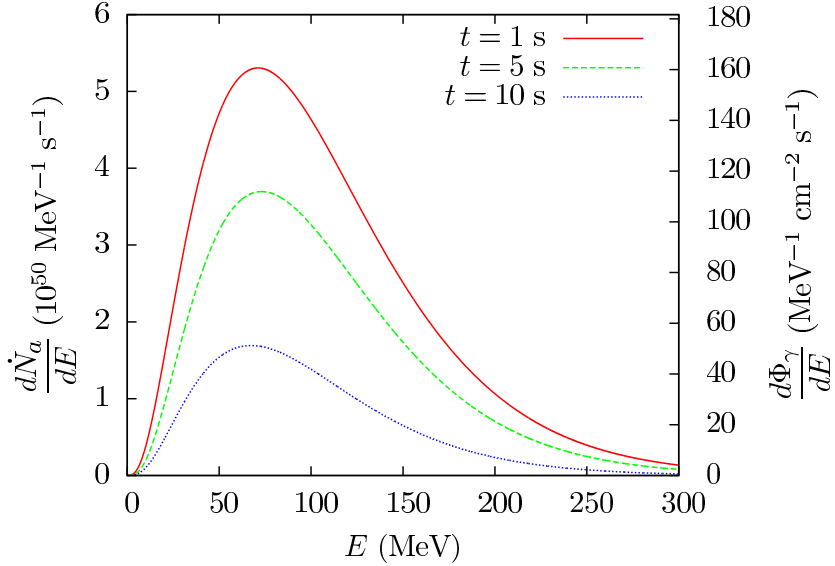}
\caption{Case of a nearly massless ALP ($m_a \lesssim 10^{-11}~$eV) with $g_{a\gamma} = 10^{-10}$~GeV$^{-1}$.
\emph{Left axis}: Total ALP production rate per unit energy for the $18$~$M_{\odot}$ progenitor including degeneracy and mass-reduction effects. \emph{Right axis}: Corresponding differential photon flux per unit energy at the Earth, for which the conversion probability was calculated using the model of Jansson and Farrar for the Galactic magnetic field in the direction of SN1987A.}
\label{fig:dNdotdE_and_dFluxgammadE_fischer_18Msun_meandd_jf}
\end{figure}

We therefore estimate the total ALP production rate per unit energy via
\begin{equation}
\frac{d\dot{N}_a}{dE} = \int_{0}^{r_{\rm max}} 4\pi r^2 \frac{d\dot{n}_a}{dE} \ dr\,\,\,,\label{eq:dNdotdE}
\end{equation}
which we show in Fig.~\ref{fig:dNdotdE_and_dFluxgammadE_fischer_18Msun_meandd_jf}, obtained by interpolating the radial dependence of the physical quantities entering Eq.~\eqref{eq:axprod} with cubic splines. For definiteness, we present the results for the $18$~$M_{\odot}$ progenitor and, to allow for a direct comparison of our findings with the ones of Ref.~\cite{Brockway:1996yr}, in Fig.~\ref{fig:dNdotdE_and_dFluxgammadE_fischer_18Msun_meandd_jf} we illustrate what we obtain using the SN simulations corresponding to three representative post-bounce times, namely 1, 5, and 10~s after the core bounce, and take a representative value of $g_{a\gamma}=10^{-10}$~GeV$^{-1}$.
	Of course, in the following, we make use of all the simulation data describing the evolution of the SN core after the bounce, thereby obtaining for the first time a detailed evolution of the ALP production with time.

An excellent fit to the total production rate is provided by the following expression~\cite{Andriamonje:2007ew},
also widely used in SN neutrino studies,
\begin{equation}
\frac{d\dot{N}_a}{dE} = C \left(\frac{E}{E_0}\right)^\beta
e^{-(\beta+1)E/E_0} \,\ .\label{eq:fit_dNdotdE}
\end{equation}
Here $C$ is a normalization constant while the fit parameter $E_0$ coincides with the average energy $\langle E_a \rangle = E_0$. Numerically, we find $C=1.03\times 10^{52}$~MeV$^{-1}$~s$^{-1}$, $E_0=\nobreak105.6 $~MeV, $\beta=\nobreak 2.145 $ for the curve corresponding to $t = 1$~s shown in Fig.~\ref{fig:dNdotdE_and_dFluxgammadE_fischer_18Msun_meandd_jf}.
Similarly, we document for future use the fit parameters for other values of the after-bounce time in Table~\ref{table:fits}. For completeness, we also calculate the time-integrated spectrum, by integrating the total ALP production rate per unit energy over the duration of the explosion. It shares the form of the instantaneous spectrum given in Eq.~\eqref{eq:fit_dNdotdE}, but with $C=9.31\times 10^{52}$~MeV$^{-1}$, $E_0=\nobreak102.3 $~MeV, $\beta= 2.25$. The total energy carried away by ALPs in this $g_{a\gamma}=10^{-10}$~GeV$^{-1}$ case is $E_{\rm tot} = 8.47\times10^{49}$~erg.

\begin{table}[h!!]
\begin{center}
\begin{tabular}{c||c|c|c}
%time after bounce
$t$ (s)	& $C$ ($10^{52}$ MeV$^{-1}$ s$^{-1}$)	& $E_0$ (MeV)	& $\beta$ \\\hline\hline
0.005	& $6.24\times10^{-2}$ 				& 35.2 		& 2.25	\\
0.2	& $3.94\times10^{-1}$				& 77.3		& 2.02	\\
0.5	& $8.05\times10^{-1}$				& 98.5		& 2.065	\\
1	& 1.03	 					& 105.6 	& 2.145	\\
1.5	& 1.1						& 107.6		& 2.19	\\
2	& 1.11						& 108.3		& 2.22	\\
3	& 1.07						& 107.8		& 2.28	\\
4	& $9.85\times10^{-1}$				& 106.7		& 2.315	\\
5	& $8.82\times10^{-1}$				& 105.3		& 2.34	\\
6	& $7.74\times10^{-1}$				& 103.9		& 2.35	\\
7	& $6.66\times10^{-1}$				& 102.4		& 2.355	\\
8	& $5.68\times10^{-1}$				& 100.8		& 2.355	\\
9	& $4.85\times10^{-1}$				& 99.4		& 2.35	\\
10	& $4.11\times10^{-1}$				& 97.5		& 2.35	\\
11	& $3.49\times10^{-1}$				& 95.8		& 2.35	\\
12	& $2.98\times10^{-1}$				& 93.7		& 2.35	\\
13	& $2.53\times10^{-1}$				& 91.6		& 2.35	\\
14	& $2.14\times10^{-1}$				& 89.5		& 2.35	\\
15	& $1.78\times10^{-1}$				& 87.2		& 2.355	\\
16	& $1.55\times10^{-1}$				& 85.8		& 2.355	\\
17	& $1.3\times10^{-1}$				& 82.9		& 2.37	\\
18	& $1.11\times10^{-1}$				& 80.2		& 2.385	\\
\end{tabular}
\end{center}
\caption{Parameters for the fit to the total ALP production rate per unit energy (in MeV$^{-1}$ s$^{-1}$) given in Eq.~\eqref{eq:fit_dNdotdE}, for a few times $t$ after the core bounce. These provide a good approximation of the numerical results obtained using the 18~$\Msun$ progenitor, including degeneracy and mass-reduction effects. The values in this table were obtained using an ALP--photon coupling $g_{a\gamma} = 10^{-10}$~GeV$^{-1}$; for other values of the coupling, the total ALP production rate per unit energy scales proportionally with $g_{a\gamma}^2$ (see Eqs.~\eqref{eq:axprod} and \eqref{eq:dNdotdE}).}
\label{table:fits}
\end{table}

Now, if we are interested in the differential ALP flux per unit energy at Earth, since the emission is necessarily isotropic in our model, we should simply consider
\begin{equation}
\frac{d\Phi_{a}}{dE}= \frac{1}{4 \pi d^2} \frac{d\dot{N}_a}{dE} \,,
\end{equation}
with $d$ the distance to the supernova, which is 50~kpc in our case ($1~{\rm kpc} = 3.086\times10^{21}$~cm).

\section{ALP--photon conversions in the Milky Way}\label{sec:conv}

Once ALPs are produced in a SN core, they can easily escape the star since their mean free path in stellar matter is sufficiently large for the values of the coupling $g_{a\gamma}$ that we are considering~\cite{Brockway:1996yr}.
Then, they will propagate until they reach the Milky Way, where they can convert into photons in the Galactic magnetic field. 
Indeed, the Lagrangian given in Eq.~\eqref{eq:axlagrang} would trigger ALP--photon oscillations in external magnetic fields.

The problem of ALP--photon conversions  simplifies  if one considers the case in which ${\bf B}$ is homogeneous. We denote by ${\bf B}_T$ the transverse magnetic field, namely its component in the plane normal to the
photon beam direction. The linear photon polarization state parallel to the transverse field direction ${\bf B}_T$ is then denoted by $A_{\parallel}$ and the orthogonal one by $A_{\perp}$. 
The  component $A_{\perp}$ decouples away,
while the probability for a photon emitted in the state $A_{\parallel}$ to oscillate into an ALP after traveling a distance $d$ (and \textit{vice versa}) is given by~\cite{Raffelt:1987im}
%........................................................................
\begin{equation}
P_{a\gamma} = (\Delta_{a \gamma} d)^2 \frac{\sin^2(\Delta_{\rm osc} d/2)}{(\Delta_{\rm osc} d/2)^2} \,\ ,
\label{conv}
\end{equation}
%..............................................................................
where the oscillation wave number is~\cite{Raffelt:1987im}
%.....................................................
\begin{equation}
\Delta_{\rm osc} \equiv \left[(\Delta_{a} - \Delta_{\rm pl})^2 + 4 \Delta_{a \gamma}^2 \right]^{1/2} \,\ ,
\end{equation}
with  $\Delta_{a\gamma} \equiv {g_{a\gamma} B_T}/{2} $ and $\Delta_a \equiv - {m_a^2}/{2E}$. The term $\Delta_{\rm pl} \equiv -{\omega^2_{\rm pl}}/{2E}$ accounts for plasma effects, where  $\omega_{\rm pl} \simeq 3.69 \times 10^{- 11} \sqrt{n_e /{\rm cm}^{- 3}} \, {\rm eV}$ is the plasma frequency expressed as a function of the electron density in the medium, $n_e$. For typical values of the relevant parameters in our Galaxy, numerically one finds
\begin{eqnarray}  
\Delta_{a\gamma}&\simeq &   1.5\times10^{-2} \left(\frac{g_{a\gamma}}{10^{-11}~\textrm{GeV}^{-1}} \right)
\left(\frac{B_T}{10^{-6}~\rm G}\right)~{\rm kpc}^{-1}
\nonumber\,,\\  
\Delta_a &\simeq &
 -7.8 \times 10^{-3} \left(\frac{m_a}{10^{-10}~{\rm eV}}\right)^2 \left(\frac{E}{100~{\rm MeV}} \right)^{-1}~{\rm kpc}^{-1}
\nonumber\,,\\  
\Delta_{\rm pl}&\simeq & 
  -1.1\times 10^{-6}\left(\frac{n_e}{10^{-3}~{\rm cm}^{-3}}\right)\left(\frac{E}{100~{\rm MeV}}\right)^{-1}~{\rm kpc}^{-1}
\nonumber\, .
\label{eq:Delta0}\end{eqnarray}
One then realizes that for $E\sim 100$~MeV, $P_{a\gamma} \simeq (\Delta_{a \gamma} d)^2$ becomes energy independent since $\Delta_{a\gamma} \gg \Delta_a, \Delta_{\rm pl}$.
 
Measurements of the Faraday rotation based on pulsar observations have shown that the regular component of  the Galactic
$B$ field  is parallel to the Galactic plane, with a typical strength $B \simeq$ a few~$\mu{\rm G}$,
with  a radial coherence length  $l_r \simeq10~{\rm kpc}$~\cite{Beck:2008ty}. Inside the Milky-Way disk the electron density is \mbox{$n_e \simeq 1.1 \times 10^{-2}~{\rm cm}^{-3}$}~\cite{Digel}, resulting in a plasma frequency ${\omega}_{\rm pl} \simeq 4.1 \times 10^{-12}~{\rm eV}$.
Among the possible $B$-field models proposed in literature, 
we take the recent Jansson and Farrar model~\cite{Jansson:2012pc} as our benchmark. 
We have also checked the model discussed by Pshirkov
\textit{et al.}\@~\cite{Pshirkov:2011um},
 which gives results similar to the one of Ref.~\cite{Jansson:2012pc} in the regions of our interest. 

Due to the presence of a rather structured behavior in the Galactic $B$ field,
 the propagation of ALPs in the Galaxy is clearly a truly 3-dimensional problem, because\hspace{1pt}---\hspace{1pt}due to the variations of the direction of ${\bf B}$\hspace{1pt}---\hspace{1pt}the same photon
polarization states play the role of either $A_{\parallel}$ and  $A_{\perp}$ in
different domains.
We have closely followed the technique described in Ref.~\cite{Horns:2012kw} (to which we address the reader for more details) to
solve the beam propagation equation along a Galactic line of sight.
In particular, the  position of SN1987A would correspond to a Galactic latitude $b=-32.1^\circ$ and longitude
$l=279.6^\circ$.
  An illustrative sky map of the line-of-sight dependent probability for an ALP at the edge of the Galaxy to convert into a photon at Earth is shown in Ref.~\cite{Horns:2012kw}   for our chosen reference magnetic field model~\cite{Jansson:2012pc}.

In Figure~\ref{fig:dNdotdE_and_dFluxgammadE_fischer_18Msun_meandd_jf}, we show the differential photon flux per unit energy arriving at Earth,
%.......................................................
\begin{equation}
\frac{d\Phi_{\gamma}}{dE}= \frac{1}{4 \pi d^2} \frac{d\dot{N}_a}{dE} \times P_{a\gamma} \,,\label{eq:diffphotonflux}
\end{equation}
%.......................................................
taking $d=50$~kpc the distance of SN1987A, and using the model of Jansson and Farrar,  for an ALP with $m_a \lesssim 10^{-11}$~eV. For such a nearly massless ALP, the conversion probability is actually essentially independent of the energy, which is why we can show this graph together with the total production rate: for the mass and coupling considered there, $P_{a\gamma} = 9\times10^{-2}$ (while the model of Pshirkov~\textit{et al.}\@ would actually give $P_{a\gamma} = 8\times10^{-1}$). For larger values of the mass, which we also consider in the following to provide the mass-dependence of our limit, the conversion probability can then become very energy dependent. 

\section{Limit on the fluence}
\label{sec:results}

\subsection{Our state-of-the-art upper limit}

At the time of the SN1987A explosion, the Gamma-Ray Spectrometer (GRS) on the Solar Maximum Mission was operative.
Even if it was pointing in the direction of the Sun, it would have been able to observe gamma rays coming from SN1987A, located at a $90^\circ$ angle with the viewing direction, and which would have therefore been seen through the shielding of the instrument~\cite{Chupp:1989kx}.\footnote{This instrument has demonstrated its ability to do so since it discovered, a few months after the explosion, the predicted gamma-ray lines associated with the decay of radioactive isotopes formed inside SN1987A, which was again at that time seen through the shielding, thereby confirming that nucleosynthesis took place~\cite{Matz:1987gl}.}
\begin{figure}
\centering
\includegraphics[width=.75\columnwidth]{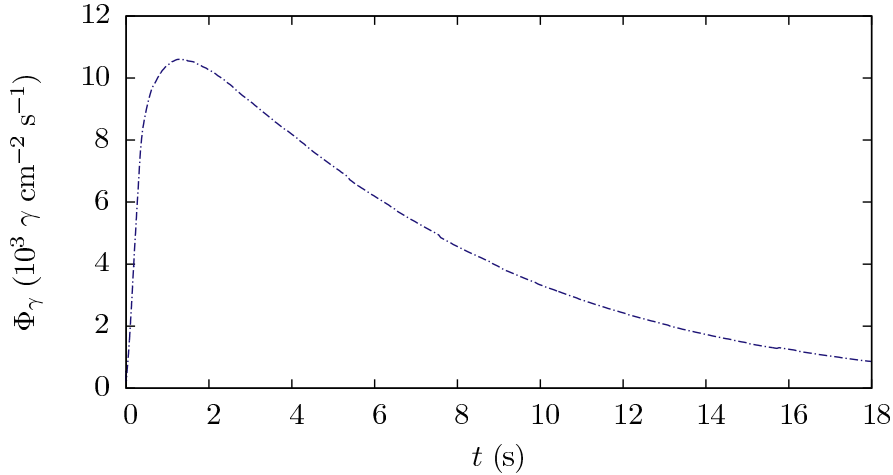}
\caption{Photon flux at Earth for photons with energy between 25--100~MeV as a function of time, for the $18$~$M_{\odot}$ progenitor including degeneracy and mass-reduction effects, using $g_{a\gamma} = 10^{-10}$~GeV$^{-1}$, $m_a \lesssim 10^{-11}$~eV, and the model of Jansson and Farrar for the Galactic magnetic field. For convenience, we have now shifted the origin of $t$ to compensate for the time it takes for an ALP produced in the supernova to reach the Earth.}
\label{fig:flux_fischer_18Msun_meandd_jf}
\end{figure}

In particular, ultralight ALPs would have led to a gamma-ray burst from SN1987A in coincidence with 
the observations
of neutrinos in the Irvine--Michigan--Brookhaven~\cite{Bionta:1987qt}, the Kamiokande-II~\cite{Hirata:1987hu}, and the Baksan~\cite{Alekseev:1987ej} experiments. 
The presence of such a signal was searched for: however, since the GRS instrument did not observe any photon excess in the 4.1--100~MeV range~\cite{Chupp:1989kx}, only 
upper bounds were placed on the fluence, which is the integral of the flux over time.

As discussed in both of the original analyses that we follow~\cite{Brockway:1996yr,Grifols:1996id} and as readily understood from Fig.~\ref{fig:dNdotdE_and_dFluxgammadE_fischer_18Msun_meandd_jf}, the best constraint on ALP--photon conversion then comes from the highest energy bin $[25,100]$~MeV, for which the fluence has been constrained to be smaller than $0.6~\gamma$~cm$^{-2}$ at 3$\sigma$ C.L. during the neutrino burst duration, namely 10.24~s~\cite{Chupp:1989kx}.

We have calculated the expected fluence from SN1987A by integrating the flux in the highest GRS energy bin, shown in Fig.~\ref{fig:flux_fischer_18Msun_meandd_jf}, over this time window, and compared this with the observational limit. Since we take into account the proton degeneracy, we consider all contributions to the ALP production right after the core bounce. 

\begin{figure}[t]
\centering
\includegraphics[width=\textwidth]{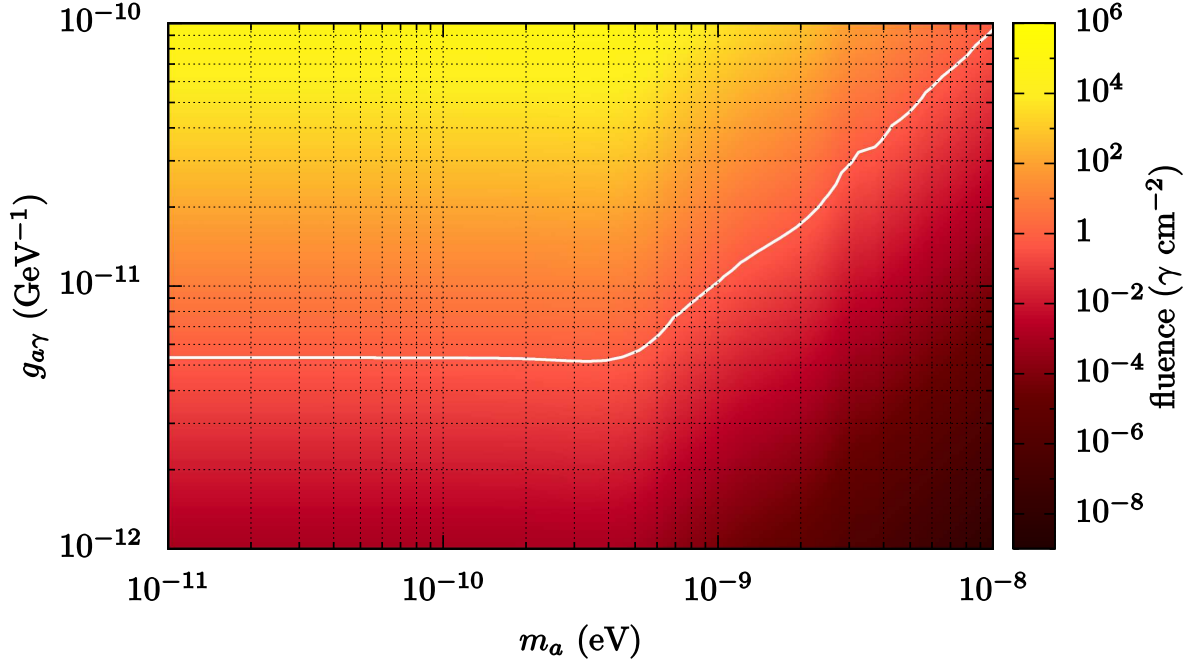}
\caption{Fluence as a function of the ALP mass $m_a$ and coupling $g_{a\gamma}$, together with our updated upper limit, shown as a white curve corresponding to $\mathrsfso{F}(m_a, g_{a\gamma}) = 0.6\ \gamma$~cm$^{-2}$. The model of Jansson and Farrar for the Galactic magnetic field is used (see text for details).}
\label{fig:bound_18Msun_JF}
\end{figure}

Of course, what we have discussed thus far was only for the case of an ALP of mass $m_a = 10^{-11}$~eV and coupling $g_{a\gamma} = 10^{-10}$~GeV$^{-1}$. To get a precise mass dependence of the limit, we actually perform a detailed scan of the ALP parameter space $(m_a, g_{a\gamma})$. Our results are shown in Fig.~\ref{fig:bound_18Msun_JF}, using the model of Jansson and Farrar, the $18$~$M_{\odot}$ progenitor, and including degeneracy and mass-reduction effects. Note that the fluence spans many orders of magnitude: a small increase in $g_{a\gamma}$ actually corresponds to a large modification of the expected fluence, which essentially goes as $g_{a\gamma}^4$ on this graph. This makes this bound very stable against various changes in the model.
In particular, for nearly massless ALPs, we obtain the limit:
\begin{equation}
g_{a\gamma} \lesssim 5.3 \times 10^{-12}\,\ \textrm{GeV}^{-1},\,\  \,\ \textrm{for}\,\ \,\ m_{a} \lesssim 4.4 \times 10^{-10}~\textrm{eV} \,\ ,\label{eq:newlimit}
\end{equation}
which has a mass dependence such that it becomes $g_{a\gamma} \lesssim 10^{-11}$~GeV$^{-1}$ at $m_{a} \simeq 10^{-9}$~eV.
Finally, since it may be useful for future reference, note that for nearly massless ALPs of mass $m_a \lesssim 10^{-11}$~eV, the fluence in this figure is very well approximated by
\begin{equation}
\mathrsfso{F}(g_{a\gamma}) = 7.02\times 10^4 {\left(\frac{g_{a\gamma}}{10^{-10}~\textrm{GeV}^{-1}}\right)}^4~\gamma~\textrm{cm}^{-2}.\label{eq:fluence_mainresult}
\end{equation}

\subsection{Discussions of the physics and robustness of our results}

The results presented thus far have been obtained using the $18$~$M_{\odot}$-progenitor supernova model, the model of Jansson and Farrar for the magnetic field, and including the effects of proton degeneracy and of the modification of the proton mass in the nuclear medium. However, we have also performed a number of comparisons using other models and their combinations. In the following subsections we investigate the stability of our bound on $g_{a\gamma}$ under various changes, and we discuss the physical implications of the new effects that were here taken into account for the first time.

\subsubsection{Degeneracy}

\begin{figure}[t]
\centering
\includegraphics[width=.75\columnwidth]{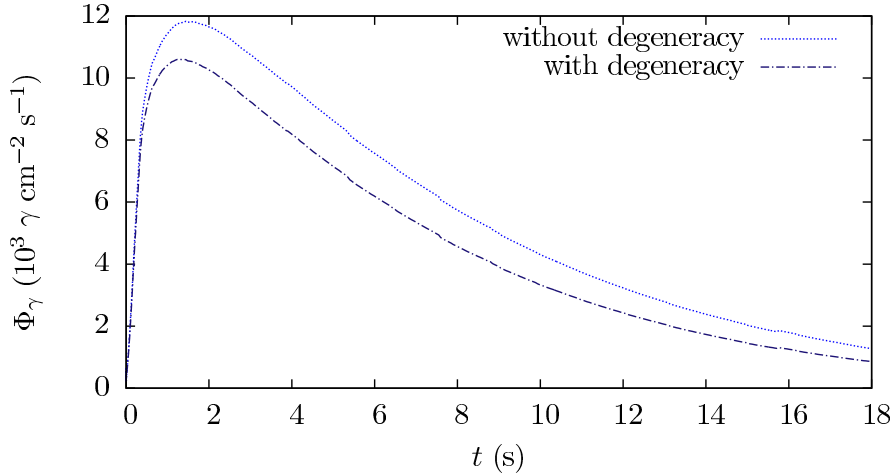}
\caption{Comparison of the photon flux for energies between 25--100~MeV, obtained using the $18$~$M_{\odot}$ progenitor and the model of Jansson and Farrar, including mass-reduction effects, without and with  (same as in Fig.~\ref{fig:flux_fischer_18Msun_meandd_jf}) degeneracy.}
\label{fig:flux_fischer_18Msun_jf_discussion}
\end{figure}
In this work we have included the effects of proton degeneracy in the Primakoff emission rate, accounting also for the reduction of the proton mass in the dense medium characteristic of the SN soon after the explosion. Lighter protons are more degenerate, and this tends to decrease the production rate. However, as protons are more easily degenerate, they form a slightly stiffer background, therefore contributing less to the screening. The combination of these effects causes a small reduction of the emission rate, as shown in Fig.~\ref{fig:flux_fischer_18Msun_jf_discussion}.

The effects of proton degeneracy were not included in the two original studies and it was argued in Ref.~\cite{Brockway:1996yr} that such effects would be mostly important during the first second after the core bounce.
What we find is that the strongest effects of degeneracy are actually not found at $t<1$~s but at later times. 
This apparently counter-intuitive result comes from the fact that ALPs are produced mostly in the hottest regions of the SN core, which do not necessarily include the central region. Indeed, as already mentioned in Sec.~\ref{sec:ALPprod} and seen in Fig.~\ref{fig:contours}, in the first instants after the bounce the temperature is higher at larger radii. Therefore, even though the core is more degenerate initially, during these early times it is not contributing substantially to the ALP production. 

\subsubsection{Stellar model}

The effect of the progenitor is in fact rather mild, which is very good for the stability of our limit. In addition to the 18~M$_\odot$ progenitor model, we calculate the ALP production for a lighter massive progenitor star with 10.8~M$_\odot$. Differences between these two stellar models, e.g.\@ in terms of peak temperatures and other nuclear matter properties relevant for the ALP production, are actually of the order of a few percent. Since the limit goes as the fourth root of the fluence, the temperature and the density should have changed much more than they do to lead to a substantial modification.

We have also compared the results obtained using the new SN models with the old model of Ref.~\cite{Brockway:1996yr}, which we have also considered in our analysis. For a given magnetic field model, with the 10.8~$\Msun$ progenitor, the limit would be weaker while, with the 18~$\Msun$ one, it is slightly more stringent than with the old model (essentially both because in the old model the maximum of production is at higher energies due to a higher temperature, and because we are not limited to the interval 1--10~s). We compare the bounds obtained using the new SN models in Fig.~\ref{fig:bound_18Msun_pshirkov}, where the effects of degeneracy and mass reduction are included in all cases.

\subsubsection{Magnetic field models}

\begin{figure} [t]
\centering
\includegraphics[width=.6\columnwidth]{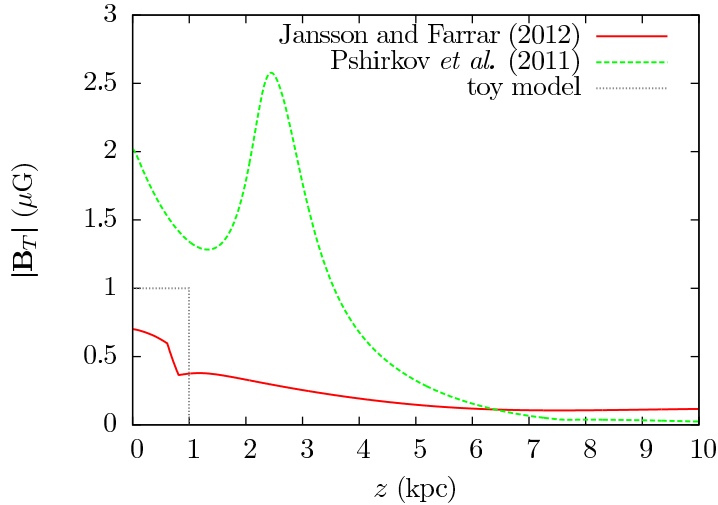}
\caption{Norm of the transverse Galactic magnetic field as a function of the distance in the direction of SN1987A in various models.}
\label{fig:BTtowardsSN1987A}
\end{figure}

The conversion probability that we find using each Galactic magnetic field model is larger than in the toy model used in Refs.~\cite{Brockway:1996yr,Grifols:1996id}, which considered a homogeneous magnetic field region of about 1~$\mu$G, over 1~kpc.\footnote{While the 10.8~$\Msun$ progenitor leads to a smaller ALP flux than the model originally considered in Ref.~\cite{Brockway:1996yr}, the larger conversion induced in the two updated Galactic magnetic field models actually leads to slightly more stringent constraints on $g_{a\gamma}$, even in that case.} The main reason for this difference, is that such a toy model ignored the presence of a halo component; see Fig.~\ref{fig:BTtowardsSN1987A}.

%%%%%%%%%%%%%%%%%%%%%%%%%%%%%%%%%%%%%%%%%%%%%%%%%%%%%%%%%%%%%%%%
%%%%%%%%%%%%%%%%%%%%%%%%%%%%%%%%%%%%%%%%%%%%%%%%%%%%%%%%%%%

Furthermore, using the model of Pshirkov~\textit{et al.}\@~\cite{Pshirkov:2011um} instead of the one of Jansson and Farrar~\cite{Jansson:2012pc} actually makes our limit change more than it does when we consider different supernova simulations. 
The model of Pshirkov~\textit{et al.}\@ in fact leads to more stringent bounds on ALPs than the one of Jansson and Farrar; the field strength is indeed significantly larger over the first 6~kpc, as seen in Fig.~\ref{fig:BTtowardsSN1987A}.

\begin{figure}[t]
\centering
\includegraphics[width=\textwidth]{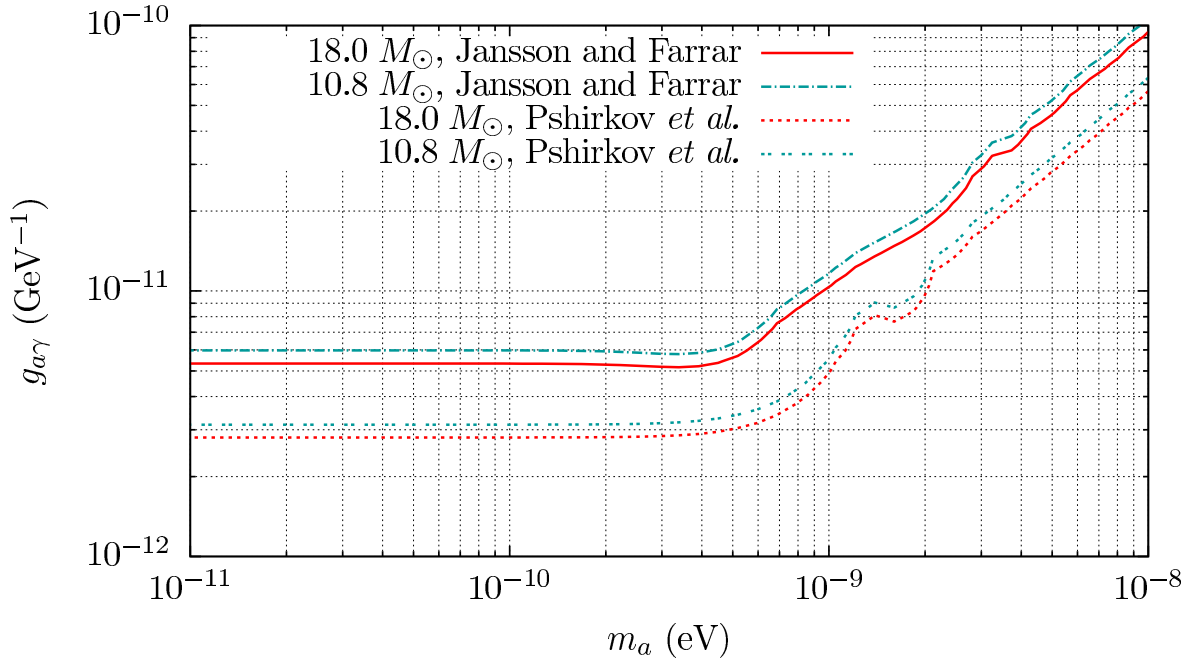}
\caption{Upper limit obtained for the 10.8~$\Msun$ and the 18~$\Msun$ progenitors, using either the model of Jansson and Farrar or the one of Pshirkov~\textit{et al.}\@ for the Galactic magnetic field.}
\label{fig:bound_18Msun_pshirkov}
\end{figure}

For completeness, we show the bound that we get with this other Galactic magnetic field model in Fig.~\ref{fig:bound_18Msun_pshirkov}. This has been obtained in the same conditions as Fig.~\ref{fig:bound_18Msun_JF}: namely, including both degeneracy and mass-reduction effects.
For the 18~$\Msun$ progenitor, we then find
%%%%%%%%%%%%%%%%%%%%%%%%%%%%%%%%%%%%%%%%%%%%%%%%%%%%%%%%%%%%%%%%%%%%%%%%%%%%
 \begin{equation}
 g_{a\gamma} \lesssim 2.8 \times 10^{-12}\,\ \textrm{GeV}^{-1} \textrm{ for } m_{a} \lesssim 3.2 \times 10^{-10}~\textrm{eV},
 \end{equation}
%%%%%%%%%%%%%%%%%%%%%%%%%%%%%%%%%%%%%%%%%%%%%%%%%%%%%%%%%%%%%%%%%%%%%%%%%%% 
 and with a mass dependence such that it becomes $g_{a\gamma} \lesssim 4.9 \times 10^{-12}$~GeV$^{-1}$ at $m_{a} \simeq 10^{-9}$~eV.

We use the bound that we obtain with the model of Jansson and Farrar as our main result since it leads to the most conservative limit.
In addition, this model provides a good fit to combined observations of Galactic synchrotron emission maps and more than 40,000 extragalactic rotation measures, which substantially improved our knowledge of the out-of-plane component.

\subsubsection{Dependence on the effective area}

The absence of any photon excess at the time of the neutrino burst from SN1987A in any of the three energy bands of the GRS instrument led to 3$\sigma$ upper limits on the observed fluence: $0.9~\gamma$~cm$^{-2}$ for 4.1--6.4~MeV; $0.4~\gamma$~cm$^{-2}$ for 10--25~MeV; $0.6~\gamma$~cm$^{-2}$ for 25--100~MeV~\cite{Chupp:1989kx}. As mentioned in this observational paper, those limits have been obtained after assuming a certain spectral shape for the photons, as needed to derive an effective area of the instrument for each energy band. In particular, the authors assumed a falling spectrum of the type $E^{-2}$ for the two energy bands above 10~MeV, and obtained 115 and 63~cm$^2$ for the effective area.

However we could not reproduce these values. There is only little information about the response of this old instrument in the literature~\cite{Forrest:1980,Cooper:1985,Forrest:1986,ManyFacesoftheSun:1996wa}, and of course mostly for photons with normal incidence with its front surface: for instance, Ref.~\cite{Forrest:1980} gives $\sim100$~cm$^{2}$ as the typical GRS effective area in the 10--100~MeV range. SN1987A was however seen perpendicularly with the viewing direction, for which we only know that the response dropped by a factor up to 3 at 20~MeV, as obtained from the Monte Carlo calibration~\cite{Cooper:1985}.

Reference~\cite{Chupp:1989kx} acknowledges a private communication, but does not give more details on how to obtain their estimated effective area $S$; they only say that they then derive the 3$\sigma$ upper limit on the observed fluence in an energy band $\Delta E$ via
\begin{equation}
\mathrsfso{F}_{\textrm{obs}}\left(\Delta E\right) \leq \left(\frac{30}{S}\right){\left(\frac{2B}{10}\right)}^{\frac{1}{2}} \gamma\, \textrm{cm}^{-2},\label{eq:fluence_dep_effarea}
\end{equation}
with $B$ the background rate, which is 17.4 and 6.3 counts s$^{-1}$ in the two higher energy bands.

This means that we could not update the argument using the exact spectral shape expected from an ALP burst: as it was also the case in the original papers~\cite{Brockway:1996yr,Grifols:1996id}, there is therefore a remaining uncertainty coming from the effective area that one should use for the case at hand. We note however that since the limit on the coupling $g_{a\gamma}$ goes as a fourth root of the fluence, it is clear from Eq.~\eqref{eq:fluence_dep_effarea} that the dependence on the effective area is rather mild.

\section{Conclusions and future perspectives}
\label{sec:ccl}

Ultralight ALPs can be efficiently produced by the Primakoff process in the hot and dense stellar medium of a core-collapse SN core and convert into photons in the magnetic field of the Milky Way, potentially producing an observable gamma-ray flux. This mechanism was used in Refs.~\cite{Brockway:1996yr,Grifols:1996id} to constrain the ALP--photon coupling $g_{a\gamma}$ from the 
absence of a gamma-ray flash detected by the Gamma-Ray Spectrometer (GRS) at the time of the SN1987A explosion. The bounds presented in these two seminal papers differ from each other by roughly one order of magnitude. They were based on meanwhile outdated SN models and on a rather schematic characterization of the ALP--photon conversions in the Milky Way. Therefore, given the relevance of the SN1987A bound for the current ALP searches, here we have improved several important aspects of the previous analyses and consequently provided a revised upper limit for the value for the ALP--photon coupling. We have calculated the SN ALP flux based on state-of-the-art SN models~\cite{Fischer:2009af,Fischer:2012}. Furthermore, we have included the effects of proton degeneracy and of proton-mass reduction in the dense stellar medium on the ALP production, based on the same microscopic nuclear model that was used in the SN simulations~\cite{Shen:1998}. For the calculation of the  ALP--photon conversion in the Galactic magnetic field, we have applied sophisticated models of the Milky-Way magnetic field~\cite{Jansson:2012pc,Pshirkov:2011um}.  As a result of these novel inputs, we obtain a stringent upper limit, shown in Fig.~\ref{fig:bound_18Msun_JF}. This bound currently represents the strongest constraint on the two-photon coupling of ultralight ALPs (with $m_a \lesssim 10^{-9}$~eV). It significantly reduces the parameter space available to explain the decreased opacity of the universe to TeV photons in terms of photon--ALP conversions. Indeed, for $m_a \lesssim 10^{-10}$~eV, one should invoke very optimistic models of the cosmic magnetic field in order to have significant conversions~\cite{Meyer:2013pny,Meyer:2014gta}.

We have studied the robustness of our bound on the ALP--photon coupling considering various physical aspects, in order to quantify how potential uncertainties could affect our results. We find that the largest uncertainty is by far due to the Galactic magnetic field. For that, we have tested two different descriptions and have obtained that the most recent, sophisticated, and constrained model actually leads to the most conservative limit.

Moreover, the $g_{a\gamma}$ coupling probed by the SN1987A is in the range of sensitivity of next-generation ALP experiments, such as ALPS II~\cite{Bahre:2013ywa}   and IAXO~\cite{Irastorza:2011gs,Armengaud:2014gea}, which will also be sensitive to much higher masses, up to $m_a \sim 10^{-4}$--$10^{-2}$~eV.

Similar to the diffuse supernova neutrino background (DSNB)~\cite{Horiuchi:2008jz}, the existence of axions and ALPs would lead to the presence of a diffuse supernova axion background (DSAB), and to a diffuse supernova ALP background (DSALPB), respectively~\cite{Raffelt:2011ft}. In a given SN explosion, the present-day limits on the axion couplings allow the total energy release asso\-ciated with their emission to be similar to the neutrino one, namely $\sim 3\times10^{53}$~erg. On the other hand, accounting for our updated limit, ALPs are not allowed to carry more than $\sim 2\times\nobreak10^{47}$~erg. Correspondingly, the DSALPB spectral flux is then very much suppressed compared to the DSAB one (while both would peak at similar energies, around 20--30~MeV). Therefore, though very light ALPs have the advantage over axions of a much more efficient conversion into photons in astrophysical magnetic fields, facilitating their detection, the predicted DSALPB flux is practically too small to allow for such detection with current technology.

The limit provided here could be improved if a close-by SN explosion were to happen during the Fermi-LAT satellite experiment operation, because of its high sensitivity to high-energy gamma rays. More precisely, in the case of a fiducial core-collapse SN at 10~kpc in the direction of the Galactic center~\cite{Mirizzi:2006xx}, taking again a 18~$\Msun$ progenitor, we estimate the fluence due to the conversion of massless ALPs to evolve as
\begin{equation}
\mathrsfso{F}(g_{a\gamma}) = 8.32\times 10^6 {\left(\frac{g_{a\gamma}}{10^{-10}~\textrm{GeV}^{-1}}\right)}^4~\gamma~\textrm{cm}^{-2}, \textrm{ for energies above } 100~\textrm{MeV}.
\end{equation}
Due to the Fermi-LAT sensitivity for short transients~\cite{Atwood:2009ez}, this means that in optimal conditions we could then probe couplings more than one order of magnitude smaller than what we exclude in the present paper. Furthermore, in the very optimistic scenario in which the star Betelgeuse would explode during the lifetime of this satellite, Fermi LAT might then be able to almost reach $g_{a\gamma} = 10^{-13}$~GeV$^{-1}$. This once again confirms the high physics potential of astrophysical observations in the search for axion-like particles.

\section*{Acknowledgments}

It is our pleasure to thank Dieter Horns for useful discussions and cross-checks concerning the implementation of the magnetic field models; A.P. also thanks Sergey Troitsky for a comment about the Fermi-LAT sensitivity.
Finally, we thank Eduard Mass\'o and Georg Raffelt for a careful reading of and various comments on the manuscript.
The work of A.M. and the work of A.P. were supported by the German Science Foundation (DFG) within the Collaborative Research Center SFB 676 ``Particles, Strings and the Early Universe.'' T.F. acknowledges support from the Narodowe Centrum Nauki (NCN) within the ``Sonata'' program under contract No. UMO-2013/11/D/ST2/02645.

\end{document}